\documentclass[twocolumn,superscriptaddress,aps,prd]{revtex4-2}

\usepackage{graphicx}
\usepackage{dcolumn}
\usepackage{bm}
\usepackage{amsmath}
\usepackage{amssymb}
\usepackage{siunitx}
\usepackage{upgreek}
\usepackage{hyperref}
\usepackage{ulem}

\begin{document}

\title{Simulating vacuum birefringence with a diffractive beam propagation code}

\author{Aim\'{e} Matheron}\email{A.Matheron@hi-jena.gsi.de}
\affiliation{Helmholtz-Institut Jena, Fr\"obelstieg 3, 07743 Jena, Germany}
\affiliation{GSI Helmholtzzentrum f\"ur Schwerionenforschung, Planckstra\ss e 1, 64291 Darmstadt, Germany}
\affiliation{Faculty of Physics and Astronomy, Friedrich-Schiller-Universit\"at Jena, 07743 Jena, Germany}
\author{Michal~\v{S}m\'{\i}d}\email{m.smid@hzdr.de}
\affiliation{Helmholtz-Zentrum Dresden-Rossendorf, Bautzner Landstra\ss e 400, 01328 Dresden, Germany}
\author{Matt Zepf}\email{m.zepf@hi-jena.gsi.de}
\affiliation{Helmholtz-Institut Jena, Fr\"obelstieg 3, 07743 Jena, Germany}
\affiliation{GSI Helmholtzzentrum f\"ur Schwerionenforschung, Planckstra\ss e 1, 64291 Darmstadt, Germany}
\affiliation{Faculty of Physics and Astronomy, Friedrich-Schiller-Universit\"at Jena, 07743 Jena, Germany}
\author{Felix Karbstein}\email{f.karbstein@hi-jena.gsi.de}
\affiliation{Helmholtz-Institut Jena, Fr\"obelstieg 3, 07743 Jena, Germany}
\affiliation{GSI Helmholtzzentrum f\"ur Schwerionenforschung, Planckstra\ss e 1, 64291 Darmstadt, Germany}
\affiliation{Faculty of Physics and Astronomy, Friedrich-Schiller-Universit\"at Jena, 07743 Jena, Germany}

\date{\today}

\begin{abstract}
Ninety years after their prediction, quantum vacuum nonlinearities in macroscopic electromagnetic fields still await a direct experimental verification in the laboratory.
A particularly promising route towards their first measurement is the collision of counter-propagating laser beams in a pump-probe type experiment. 
Here, the key challenge is to separate the small quantum vacuum signal at the oscillation frequency of the probe that is mainly emitted in the vicinity of its forward cone from the large probe background.
While quantitatively accurate predictions of the associated quantum vacuum signals are available, to date there is no framework that combines these predictions with a diffractive beam propagation code. Such codes are designed to holistically model optical experiments and can reliably account for diffraction and absorption losses of optical devices, like lenses and apertures. The latter inevitably influence and modify both the induced signal and background components prior to their detection in experiment. The present work addresses this topical issue and reports on the first implementation of a quantum vacuum signals emission module in an established diffractive beam propagation toolkit designed for the realistic modelling of optical experiments. 
\end{abstract}

\maketitle

\section{Introduction}

Despite the impressive successes of Maxwell's electrodynamics in the accurate description of the physics of macroscopic electromagnetic fields, the precision of this classical theory will ultimately be limited by quantum corrections. Already about 90 years ago Heisenberg and his students Euler and Kockel determined the leading quantum corrections in the absence of real charges in the low-energy limit \cite{Euler:1935zz,Euler:1935qgl,Heisenberg:1936nmg}.
These effectively emerge from integrating out the charged particle fields of the underlying microscopic theory of quantum electrodynamics (QED) and supplement free Maxwell theory described by the Lagrangian ${\cal L}_{\rm M}=\frac{1}{4}F_{\mu\nu}F^{\mu\nu}$ with an interaction Lagrangian ${\cal L}_{\rm int}$ comprising infinitely many local nonlinear interactions of the prescribed electromagnetic field $F^{\mu\nu}$ in vacuum.

Charge conjugation invariance of the QED vacuum \cite{Furry:1937zz} implies that all effective interactions that are generated are even in $F^{\mu\nu}\sim\{E_i,B_i\}$.
Its electric $E_i$ and magnetic $B_i$ field components are rendered dimensionless by the {\it critical} fields $E_{\rm cr}=m^2c^3/(e\hbar)\simeq1.3\times10^{18}\,{\rm V}/{\rm m}$ and $B_{\rm cr}=E_{\rm cr}/c\simeq4\times10^9\,{\rm T}$, respectively, surpassing the strengths of all macroscopic electromagnetic fields currently available in experiment by orders of magnitude.
Hence, the latter qualify as {\it weak} fields and it is typically sufficient to consider only the effective interaction terms that scale quartic with $F^{\mu\nu}$; those scaling as $\sim(F^{\mu\nu})^{2(n+2)}$ with $n\in\mathbb{N}$ are relatively suppressed by $2n$ powers of $\{E_i/E_{\rm cr},B/B_{\rm cr}\}\ll1$. At the same time, derivative corrections \cite{Gusynin:1995bc,Gusynin:1998bt,Karbstein:2021obd} are subleading at low energies by definition and can in particular be safely neglected as long as the typical spatio-temporal scales of variation $\lambda$ of the fields fulfil $\lambda\gg\lambdabar_{\rm C}$ with reduced Compton wavelength of the electron $\lambdabar_{\rm C}=\hbar/(mc)\simeq3.8\times10^{-13}\,{\rm m}$. We emphasize that the electromagnetic fields studied in the remainder of this article perfectly meet these weak-field and low-energy conditions.
Therefore, we have ($c=\hbar=\epsilon_0=1$)
\begin{equation}
 {\cal L}_{\rm int}\simeq\frac{m^4}{5760\pi^2}\Bigl(\frac{e}{m^2}\Bigr)^4\Bigl[a\bigl(F_{\mu\nu}F^{\mu\nu}\bigr)^2 + b\bigl(F_{\mu\nu}{}^\star\!F^{\mu\nu}\bigr)^2\Bigr], \label{eq:Lint}
\end{equation}
where $a$, $b$ are low-energy constants that have a series expansion in $\alpha=e^2/(4\pi)\simeq1/137$, and ${}^\star\!F_{\mu\nu}$ is the dual field strength tensor.
At leading loop order, $a=4$ and $b=7$ \cite{Euler:1935zz}.

\begin{figure}[h!] 
  \centering
  \includegraphics[width=8.5cm]{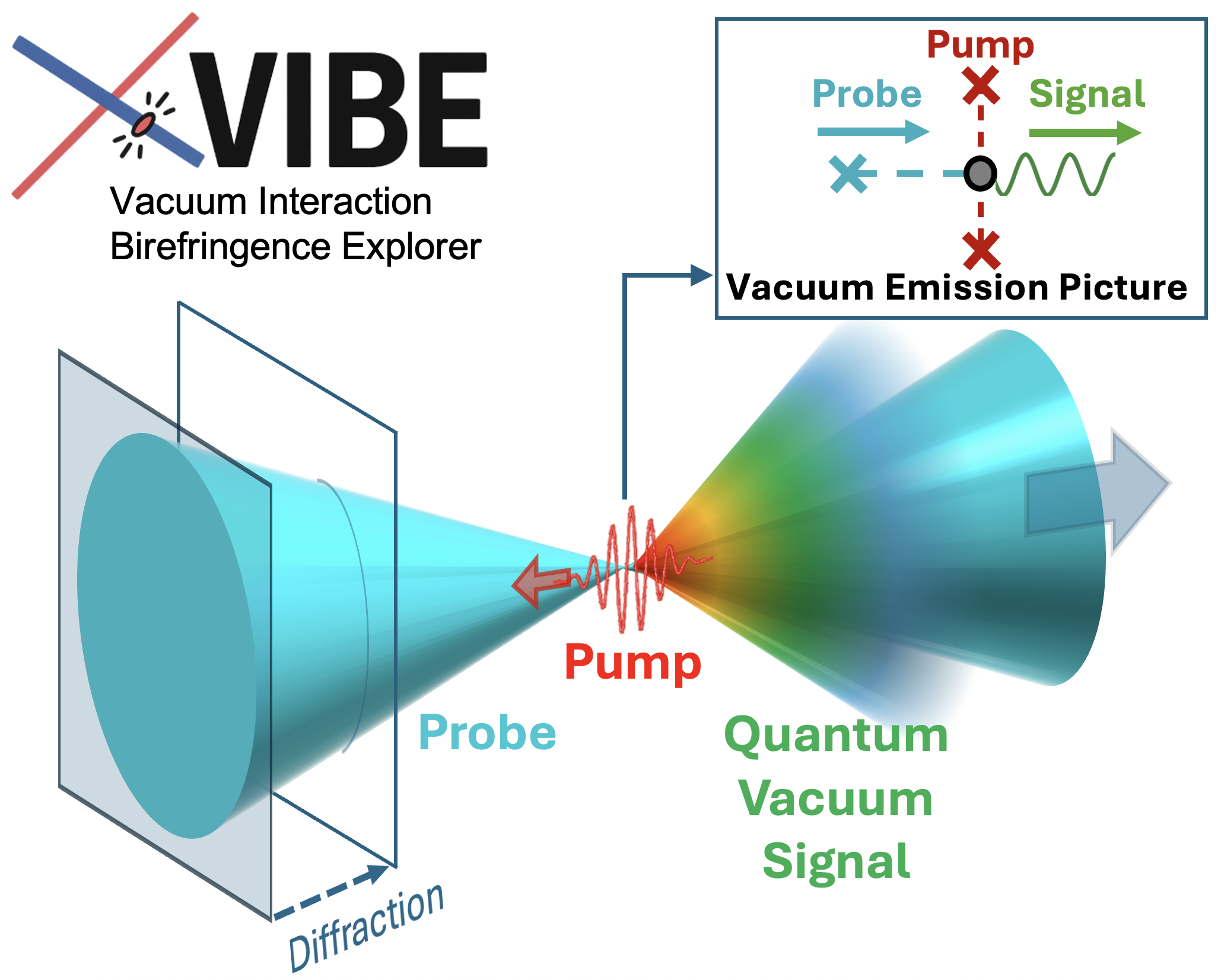}
  \caption{Schematic illustration of the vacuum emission module VIBE implemented in the diffractive beam propagation toolbox LightPipes.}
  \label{fig:1}
\end{figure}

Many proposals aiming at the first direct verification of the quantum vacuum nonlinearity predicted by Eq.~\eqref{eq:Lint} in macroscopic electromagnetic fields are available; see \cite{DiPiazza:2011tq,Battesti:2012hf,King:2015tba,Battesti:2018bgc,Karbstein:2019oej,Fedotov:2022ely} and references therein.
With regard to an experimental realization, schemes envisioning the collision of just two laser pulses in a pump-probe type experiment look particularly promising.
In this context, special attention has been put on the scenario where a brilliant x-ray probe generated by an x-ray free-electron laser (XFEL) is collided with a tightly focussed near-infrared (IR) pump laser in a counter-propagating geometry; see in particular \cite{Ahmadiniaz:2024xob} and references therein.
So far, most of the proposals for discovery experiments of quantum vacuum nonlinearity have been put forward on the basis of a simplified analytical modelling of the colliding laser fields.

While numerical quantum vacuum solvers have become available \cite{Blinne:2018nbd,Grismayer:2016cqp,Lindner:2021krv,Zhang:2025,Alawashra:2025trj}, these have mainly been employed to test and verify analytical predictions based on simplified laser pulse models and to efficiently explore parameter regimes challenging analytical approaches.
To the best of our knowledge, especially no successes towards integrating and implementing the quantum vacuum signal induced in laser pulse collision in a diffractive beam propagation code have been reported. Such codes are designed to holistically model coherent optical devices, such as lenses and apertures. Achieving this is the key idea of the present work which introduces the vacuum emission module {\it Vacuum Interaction Birefringence Explorer} (VIBE) allowing to determine the dominant quantum vacuum signal at the oscillation frequency of the probe within the diffractive beam propagation toolbox LightPipes \cite{LightPipes:1997,LightPipes:2.1.5}; see Fig.~\ref{fig:1}.
In its current version, VIBE should allow for the reliable study of this signal for the specific case of a weakly focussed probe counter-propagating a tightly focussed pump in unprecedented detail. We emphasize that these conditions are specifically met by the vacuum birefringence experiment planned to be set up at the High Energy Density (HED) scientific instrument of the European XFEL within the BIREF@HIBEF collaboration \cite{Ahmadiniaz:2024xob,Smid:2025wmj}
using the XFEL as probe and the near-IR laser ReLaX \cite{ReLaX} as pump.
For the accurate planing and analysis of this type of experiments, a formalism capable of describing experimentally realistic focus profiles of the colliding laser beams at moderate numerical costs is especially timely.
The implementation in a diffractive beam propagation code comes with additional benefits because the versatility of the former allows to consistently propagate the induced quantum vacuum signal through a complete model of the actual experimental setup.

Our present work is organized as follows. In Section~\ref{sec:Form+Impl} we briefly introduce the formalism used by us and detail the various approximations and assumptions invoked by us to arrive at tractable expressions to be straightforwardly set up and implemented in the beam propagation toolbox LightPipes. The subsequent Section~\ref{sec:Results} showcases several specific example results in the context of the recent activities at HED-HIBEF \cite{Ahmadiniaz:2024xob,Smid:2025wmj}. Finally, we end with conclusions and an outlook in Section~\ref{sec:Concls+Outlook}.

\section{Formalism and Implementation}\label{sec:Form+Impl}

Since the aim is to obtain nonlinear quantum vacuum signals by applying strong macroscopic electromagnetic fields, we decompose $F^{\mu\nu}\to F^{\mu\nu} + f^{\mu\nu}$ in Eq.~\eqref{eq:Lint} into an applied $F^{\mu\nu}$ and a signal $f^{\mu\nu}=\partial^\mu a^\nu-\partial^\nu a^\mu$ component fulfilling $\lVert f^{\mu\nu}\rVert\ll\lVert F^{\mu\nu}\rVert$, where $a^\mu$ is the {\it signal photon} field and $\lVert\cdot\rVert$ denotes an appropriate norm \cite{Bialynicka-Birula:1970nlh,Brezin:1971nd}.
Linearizing in $f^{\mu\nu}$ and using integration by parts, the four-vector
\begin{equation}
 j_\nu(x)=2\partial^\mu\frac{\partial{\cal L}_{\rm int}}{\partial F^{\mu\nu}}\,, \label{eq:j(x)}
\end{equation}
plays the role of the source of the nonlinear quantum vacuum signal.
Using standard Green's function methods, it can readily be shown that outside the interaction region and at asymptotic times the electric field vector of the induced signal can be compactly expressed as \cite{Karbstein:2019oej}
\begin{equation}
 \vec{e}_{\rm S}(x)={\rm Re}\int\frac{{\rm d}^3k}{(2\pi)^3}\,{\rm e}^{{\rm i}|\vec{k}|(\hat{\vec{k}}\cdot\vec{x}-t)}\vec{J}(\vec{k})\,, \label{eq:eS}
\end{equation}
with $\vec{J}(\vec{k}):=\bigl[\vec{j}(k)-\hat{\vec{k}}j^0(k)\bigr]\big|_{k^0=|\vec{k}|}$, where $j_\nu(k)=\int{\rm d}^4x\,{\rm e}^{-{\rm i}kx} j_\nu(x)=2{\rm i}k^\mu\int{\rm d}^4x\,{\rm e}^{-{\rm i}kx}\,\partial{\cal L}_{\rm int}/\partial F^{\mu\nu}$ is the Fourier transform of Eq.~\eqref{eq:j(x)} to momentum space and $\hat{\vec{k}}=\vec{k}/|\vec{k}|$ is a unit vector.

Now, the idea is to induce a sizeable signal component by using the strong macroscopic electromagnetic fields generated by lasers.
Because both $F_{\mu\nu}F^{\mu\nu}=2(\vec{B}^2-\vec{E}^2)$ and $F_{\mu\nu}{}^\star\!F^{\mu\nu}=-4\vec{B}\cdot\vec{E}$ tend to be very small for a single laser field (they vanish identically for a leading-order paraxial beam), this typically requires the collision of at least two laser fields.
As noted already in the introduction, a particularly promising scenario is the head-on collision of counter-propagating x-ray free-electron and near-IR high-intensity laser pulses \cite{Ahmadiniaz:2024xob}, which is also the example studied in the present work.
To be specific, we are interested in the x-ray signal at the oscillation frequency of the XFEL, constituting the dominant quantum vacuum signal in the x-ray domain for this beam configuration \cite{Karbstein:2016lby}.
In turn, we further decompose the electromagnetic field driving the effect in Eq.~\eqref{eq:j(x)} as $F^{\mu\nu}\to F^{\mu\nu}+F^{\mu\nu}_{\rm probe}$ into the fields of the near-IR {\it pump} $F^{\mu\nu}$ and the x-ray {\it probe} $F^{\mu\nu}_{\rm probe}$, and focus on the signal component linear in $F^{\mu\nu}_{\rm probe}$.
For the current in momentum space, this implies
\begin{align}
 j_\nu(k)
 &\to\,2{\rm i}k^\mu
   \int{\rm d}^4x\,{\rm e}^{-{\rm i}kx}\,
   \frac{\partial^2{\cal L}_{\rm int}}
        {\partial F^{\mu\nu}\partial F^{\rho\sigma}}\,
   F^{\rho\sigma}_{\rm probe}\,,
 \label{eq:j(x)1}
\intertext{which, for a (leading-order) paraxial pump beam and upon insertion of Eq.~\eqref{eq:Lint}, simplifies to \cite{Karbstein:2015cpa}}
 &=
 {\rm i}\frac{1}{90}\frac{\alpha}{\pi}
 \Bigl(\frac{e}{m^2}\Bigr)^2
 k^\mu \int{\rm d}^4x\,{\rm e}^{-{\rm i}kx}\,
 \notag
\\
 &\hspace*{1.6cm}\times
 \bigl(
     a\,F_{\mu\nu}F_{\rho\sigma}
     + b\,{}^\star\!F_{\mu\nu}\,{}^\star\!F_{\rho\sigma}
 \bigr)\,
 F^{\rho\sigma}_{\rm probe}\,.
 \label{eq:j(x)2}
\end{align}

Subsequently, we model pump and probe as linear polarized (leading-order) paraxial beams and focus explicitly on the counter-propagating pump-probe scenario.
Without loss of generality, the probe propagates in positive z direction. Hence, we have $-\vec{e}_{\rm z}\times\vec{E}=\vec{B}$, $\vec{e}_{\rm z}\times\vec{E}_{\rm probe}=\vec{B}_{\rm probe}$ and none of these fields has a component along $\vec{e}_{\rm z}$.
For convenience, we moreover introduce field profiles and polarization vectors for the beams as $\vec{E}={\cal E}\hat{\vec{E}}$, $\vec{E}_{\rm probe}={\cal E}_{\rm probe}\hat{\vec{E}}_{\rm probe}$, and denote the relative polarization of pump and probe by $\phi=\angle(\vec{E},\vec{E}_{\rm probe})$; the sign of $\phi$ is fixed by assuming positive orientation of the set ($\vec{E}$, $\vec{E}_{\rm probe}$, $\vec{e}_{\rm z}$).
Using Eqs.~\eqref{eq:j(x)1} and \eqref{eq:j(x)2}, the integrand in Eq.~\eqref{eq:eS} can then be expressed as
\begin{multline}
 \vec{J}(\vec{k})
 \to
 {\rm i}|\vec{k}|\frac{2}{45}\frac{\alpha}{\pi} \Bigl(\frac{e}{m^2}\Bigr)^2 \int{\rm d}^4x\, {\rm e}^{-{\rm i}|\vec{k}|(\hat{\vec{k}}\cdot\vec{x}-t)}\,{\cal E}^2{\cal E}_{\rm probe}
\\
 \times\bigl[
   a\cos\phi\,(\hat{\vec{k}}\times\hat{\vec{B}}+\hat{\vec{E}}_\perp) + b\sin\phi\,(\hat{\vec{k}}\times\hat{\vec{E}}-\hat{\vec{B}}_\perp)
 \bigr]\,,
 \label{eq:J(k)}
\end{multline}
where $\vec{v}_\perp:=\vec{v}-\hat{\vec{k}}(\hat{\vec{k}}\cdot\vec{v})$ is the component of the vector $\vec{v}$ in the direction perpendicular to $\vec{k}$.
Especially for linear polarized beams the polarization vectors of the fields and $\phi$ do not depend on space and time, such that the evaluation of Eq.~\eqref{eq:J(k)} boils down to performing the four-dimensional Fourier integral over ${\cal E}^2{\cal E}_{\rm probe}$. Being exclusively interested in the dominant quantum vacuum signal at the frequency of the probe, this can be further simplified by replacing ${\cal E}^2\to \langle{\cal E}^2\rangle_t=I$, where $\langle\cdot\rangle_t$ denotes time averaging over one oscillation period and $I$ is the intensity profile of the pump \cite{Karbstein:2016lby}.
Focusing on linear polarized beams, we now consider a monochromatic probe and a pump with Gaussian temporal profile (FWHM pulse duration $\cal T$).
For simplicity, we moreover assume both beams to be well-characterized by fixed transverse profiles throughout the entire interaction region which is confined to the vicinity of ${\rm z}=0$; without loss of generality, we define the optimal collision point to be located at ${\rm x}={\rm y}={\rm z}=0$.
In turn, we have
\begin{align}
 {\cal E}_{\rm probe}
 &= {\cal E}_{\rm probe}({\rm x},{\rm y})
   \sin\bigl(\omega_{\rm probe}({\rm z}-t)\bigr) \nonumber
\intertext{and}
 {\cal E}^2
 &\to I({\rm x},{\rm y})\,
    {\rm e}^{-\ln 2\,\bigl(\tfrac{{\rm z}+t}{{\cal T}/2}\bigr)^2}.
 \label{eq:profiles}
\end{align}
The above simplifying assumptions effectively amount to using an infinite Rayleigh range approximation (IRRA) for each of the beams; cf., e.g., \cite{Gies:2017ygp,King:2018wtn}.
In this approximation the Rayleigh range of a beam is formally sent to infinity, but its waist parameter is kept finite.
For pump and probe waists of the same order, the IRRA is usually well-justified for the x-ray probe because its Rayleigh range ${\rm z}_{\rm R,probe}$ surpasses all other length and time scales characterizing the field profiles in the interaction region by orders of magnitude \cite{Karbstein:2016lby}.
At the same time, a {\it naive} IRRA is in general not appropriate for the counter-propagating focussed near-IR pump that typically widens on lengths comparable to other characteristic scales in the interaction region.
However, specifically for the case of an IRRA probe colliding with a focussed pump featuring a Gaussian-like central peak in its transverse focus profile and being well-characterized by a Lorentzian longitudinal intensity profile $\sim 1/[1+({\rm z}/{\rm z}_{\rm R})^2]$ parametrized by a Rayleigh range parameter ${\rm z}_{\rm R}$, a procedure is available that allows to benefit from the simplifications inherent to an IRRA also for the pump \cite{Karbstein:2018omb,Mosman:2021vua}. This effectively requires supplementing the IRRA for the pump in  Eq.~\eqref{eq:profiles} with an overall rescaling factor of
\begin{equation}
 \sqrt{\tilde{F}\bigl(\tfrac{2{\rm z}_{\rm R}}{\cal T},\tfrac{t_0}{\cal T},\tfrac{{\cal T}_{\rm probe}}{\cal T}\bigr)} \label{eq:rescaling}
\end{equation}
to be accounted for in Eq.~\eqref{eq:J(k)}. The function $\tilde{F}$ appearing here is defined as
\begin{multline}
\tilde{F}(\tilde\chi,\tilde\chi_0,\rho)
 = \sqrt{\frac{\pi}{2}}\frac{\ln^{3}2}{1+2\rho^2}\,\tilde\chi^2
   \int_{-\infty}^\infty {\rm d}K\,
   {\rm e}^{-2\ln^{2}2\,K^2}\\
\times
\biggl|
  \sum_{s=\pm1}
  {\rm erfcx}\!\biggl(
    \frac{\sqrt{2}\ln 2}{\sqrt{1+2\rho^2}}
    \bigl[
      s (\rho K -{\rm i}\tilde\chi_0)
      + \tilde\chi
    \bigr]
  \biggr)
\biggr|^2,
\label{eq:tildeF}
\end{multline}
where ${\rm erfcx}(\cdot)$ is the scaled complementary error function.\footnote{Note that the function $\tilde F\equiv\tilde{F}(2{\rm z}_{\rm R}/{\cal T},t_0/{\cal T},{\cal T}_{\rm probe}/{\cal T})$ can be identically rewritten as $\tilde{F}=F/(F|_{{\rm z_{\rm R}\to\infty}})=\frac{\sqrt{3\pi}}{4}F$ with $F$ introduced in Eq.~(7) of \cite{Karbstein:2018omb}; this reference uses $1/{\rm e}^2$ pulse durations w.r.t. intensity.}
In this context the IRRA waist parameter of the pump is to be interpreted as an {\it effective waist} $w_{\rm eff}$ that is fixed such that the quantum vacuum signal induced in strict forward direction matches its exact counterpart determined without the use of an IRRA. This approximation can achieve quantitatively accurate results on the one-percent level \cite{Mosman:2021vua}.
If the condition ${\rm min}\{{\cal T},{\cal T}_{\rm probe}\}\lesssim2{\rm z}_{\rm R}$ on the FWHM pump and probe pulse durations is met and the temporal offset $t_0$ between the colliding pulses in the collision point fulfils $|t_0|\lesssim{\rm z}_{\rm R}/c$,  without a substantial loss in accuracy one may even identify $w_{\rm eff}$ with physical waist $w_0$ of the pump ($1/{\rm e}^2$ with respect to intensity).
Note that upon identifying ${\rm z}_{\rm R}$ with its Gaussian beam value ${\rm z}_{\rm R}=\pi w_0^2/\lambda$, for a representative value of ${\rm min}\{{\cal T},{\cal T}_{\rm probe}\}=25\,{\rm fs}$ and a typical near-IR wavelength of $\lambda=800\,{\rm nm}$ the former criterion translates into the condition $w_0\gtrsim1\upmu{\rm m}$ that is usually met in experiment.
Hence, here we use Eq.~\eqref{eq:profiles} together with $w_{\rm eff}=w_0$ and multiply Eq.~\eqref{eq:J(k)} by Eq.~\eqref{eq:rescaling}.

Switching to spherical momentum coordinates $\vec{k}={\rm k}(\cos\varphi\sin\vartheta,\sin\varphi\sin\vartheta,\cos\vartheta)$ and neglecting an exponentially suppressed term, the relevant Fourier integral over ${\cal E}^2{\cal E}_{\rm probe}$  can then be represented as
\begin{widetext}
\begin{multline}
  \int{\rm d}^4x\,{\rm e}^{-{\rm i}|\vec{k}|(\hat{\vec{k}}\cdot\vec{x}-t)} {\cal E}^2{\cal E}_{\rm probe}\,\to\,\delta\Bigl({\rm k}-\omega_{\rm probe}\frac{2}{1+\cos\vartheta}\Bigr)\frac{2}{1+\cos\vartheta}\,\frac{\pi^{\frac{3}{2}}}{\rm i}\frac{\cal T}{4\sqrt{\ln2}}\,{\rm e}^{-(\frac{{\cal T}}{4\sqrt{\ln2}})^2({\rm k}-\omega_{\rm probe})^2} \\
  \times\iint{\rm d}\tilde{\rm x}\,{\rm d}\tilde{\rm y}\,{\rm e}^{-{\rm ik}\sin\vartheta(\cos\varphi\,\tilde{\rm x}+\sin\varphi\,\tilde{\rm y})} \,I(\tilde{\rm x},\tilde{\rm y})\,{\cal E}_{\rm probe}(\tilde{\rm x},\tilde{\rm y})\,. \label{eq:FT-E2E}
\end{multline}
On the other hand, keeping terms up to quadratic order in $\vartheta$ in the phase of the Fourier transform in Eq.~\eqref{eq:eS}, we obtain
\begin{equation}
 \vec{e}_{\rm S}(x)\simeq{\rm Re}\int\frac{{\rm d}^3k}{(2\pi)^3}\,{\rm e}^{{\rm i}{\rm k}({\rm z}-t)}{\rm e}^{-{\rm i}\frac{1}{2}{\rm k}\vartheta^2{\rm z}+{\rm i}{\rm k}\vartheta(\cos\varphi\,{\rm x}+\sin\varphi\,{\rm y})}\vec{J}(\vec{k})\,. \label{eq:eS1}
\end{equation}
Here, the first term in the exponential ensures that for ${\rm zk}\gg1$, which is precisely the regime we are interested in (see below), the integral receives substantial contributions only from $\vartheta\ll1$.
Correspondingly, it also amounts to an excellent approximation to formally extend the integration over $\vartheta$ over all real positive values of $\vartheta$.
Making use of this, moreover replace the trigonometric functions of $\vartheta$ in Eqs.~\eqref{eq:J(k)} and \eqref{eq:FT-E2E} by their leading terms. Upon insertion of the resulting expressions into Eq.~\eqref{eq:eS1}, accounting for the rescaling factor~\eqref{eq:rescaling} and performing the $\rm k$, $\varphi$ and $\vartheta$ integrations, we then arrive at
\begin{multline}
 \vec{e}_{\rm S}(x)\simeq2\bigl(a\cos\phi\,\hat{\vec{E}} -b\,\sin\phi\, \hat{\vec{B}}\,\bigr)\sqrt{\frac{\pi}{\ln2}\,\tilde{F}\bigl(\tfrac{2{\rm z}_{\rm R}}{\cal T},\tfrac{t_0}{\cal T},\tfrac{{\cal T}_{\rm probe}}{\cal T}\bigr)}\,\frac{1}{180}\frac{\alpha}{\pi}\,{\cal T}\omega_{\rm probe} \\
 \times\,{\rm Re}\biggl\{{\rm e}^{{\rm i}\omega_{\rm probe}({\rm z}-t)}\,\frac{\omega_{\rm probe}}{2\pi{\rm i}{\rm z}}\iint{\rm d}\tilde{\rm x}\,{\rm d}\tilde{\rm y}\,{\rm e}^{{\rm i}\omega_{\rm probe}\frac{({\rm x}-\tilde{\rm x})^2+({\rm y}-\tilde{\rm y})^2}{2{\rm z}}}\,\frac{I(\tilde{\rm x},\tilde{\rm y})}{I_{\rm cr}}\,{\cal E}_{\rm probe}(\tilde{\rm x},\tilde{\rm y})\biggr\}\,, \label{eq:eS2}
\end{multline}
\end{widetext}
with critical intensity $I_{\rm cr}=c\epsilon_0E_{\rm cr}^2\simeq4.7\times10^{29}\,{\rm W}/{\rm cm}^2$.
Equation~\eqref{eq:eS2} constitutes a key result of our present work: it provides a concise expression for the electric field of the dominant quantum vacuum signal at the probe frequency induced in the head-on collision of the probe with a counter-propagating pump far outside the interaction region at ${\rm z}\gg{\rm z}_{\rm R,probe}$.
The signal field components polarized parallel ($\parallel$) and perpendicular ($\perp$) to the probe follow readily from Eq.~\eqref{eq:eS2} by contraction with $\hat{\vec{E}}_{\rm probe}$ and $\hat{\vec{B}}_{\rm probe}$, respectively. The relevant overlap factors
\begin{align}
 \hat{\vec{E}}_{\rm probe}\cdot2\bigl(a\cos\phi\,\hat{\vec{E}} -b\,\sin\phi\, \hat{\vec{B}}\,\bigr)&=c_+-c_-\cos(2\phi)\,, \nonumber\\
 \hat{\vec{B}}_{\rm probe}\cdot2\bigl(a\cos\phi\,\hat{\vec{E}} -b\,\sin\phi\, \hat{\vec{B}}\,\bigr)&=c_-\sin(2\phi)\,,
\end{align}
with $c_\pm=(b\pm a)$, are fully determined by the relative polarization of the colliding laser beams and the low-energy constants in the interaction Lagrangian~\eqref{eq:Lint}.
The $\perp$-polarized signal component can be attributed to a birefringence property of the quantum vacuum in the presence of a prescribed electromagnetic field and thus constitutes the signal of {\it vacuum birefringence} \cite{Toll:1952rq,Klein:1964znn,Baier:1967zza,Baier:1967zzc}, which has been advocated as a prospective quantum vacuum signature accessible in strong laser fields already for a long time \cite{Aleksandrov:1985,Heinzl:2006xc,DiPiazza:2006pr,King:2010kvw,Dinu:2013gaa,Dinu:2014tsa,Karbstein:2015xra,Schlenvoigt:2016jrd,Shen:2018lbq,Shen:2020,Ahmadiniaz:2022nrv,Heinzl:2024cia,Ataman:2025}.
For vacuum birefringence searches in quasi-static magnetic fields, see \cite{Battesti:2012hf,Battesti:2018bgc,Ejlli:2020yhk} and references therein.

A comparison with Eq.~(4) of the LightPipes Manual \cite{LightPipes:2.1.5} unveils that Eq.~\eqref{eq:eS2} closely resembles the Fresnel-Kirchhoff diffraction integral implemented in LightPipes to efficiently evaluate free-space propagation of a monochromatic light field of oscillation frequency $\omega_{\rm probe}$ with given initial electric field distribution ${\cal E}({\rm x},{\rm y},{\rm z}=0)$ over a distance $\rm z$ via fast Fourier transform (FFT). In our conventions, the expression given there reads
\begin{multline}
 {\cal E}({\rm x},{\rm y},{\rm z})=\frac{\omega_{\rm probe}}{2\pi{\rm i}{\rm z}}\iint{\rm d}\tilde{\rm x}\,{\rm d}\tilde{\rm y}\,{\rm e}^{{\rm i}\omega_{\rm probe}\frac{({\rm x}-\tilde{\rm x})^2+({\rm y}-\tilde{\rm y})^2}{2{\rm z}}}\\
 \times{\cal E}(\tilde{\rm x},\tilde{\rm y},0)\,. \label{eq:freespaceprop}
\end{multline}
Note that LightPipes uses a complex representation of the electric field.
Here, we make use of this analogy to determine the dominant quantum vacuum signal at the probe frequency. To this end, we introduce {\it virtual} focus profiles for the $\parallel$ and $\perp$ polarized signal components as follows,
\begin{figure*}[t!]
  \centering
  \includegraphics[width=\textwidth]{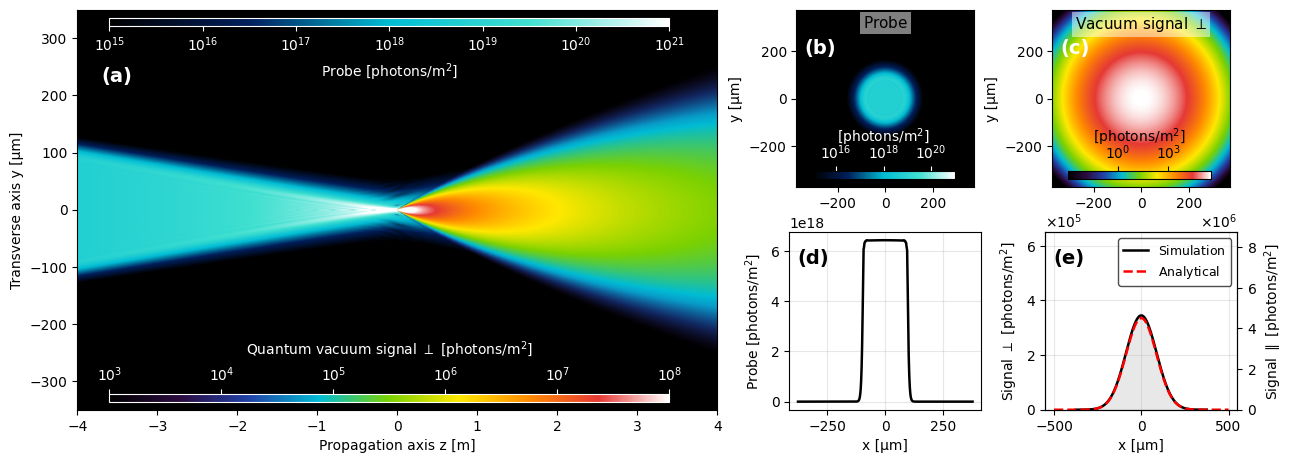}
  \caption{Idealized example scenario of a pump-probe interaction generating a quantum vacuum signal. (a) Side view of the probe (blue) and quantum-vacuum signal (multicolour) intensity maps. The signal is generated in the interaction region at ${\rm z}=0$ and co-propagates with the probe towards ${\rm z}>0$. Transverse profiles of the (b) probe and (c) signal photon distributions in the detection plane at ${\rm z}=4\,{\rm m}$. (d) Lineout through the peak value of the signal photon distribution. The black solid line is extracted from the simulation outcome (c), and the red-dotted line is the result of an analytical benchmark calculation.}
  \label{fig:2}
\end{figure*}
\begin{align}
\left\{\!\!\begin{array}{c}
 {\cal E}_{{\rm S},\parallel}({\rm x},{\rm y},0) \\
 {\cal E}_{{\rm S},\perp}({\rm x},{\rm y},0)
\end{array}\!\!\right\}
=&\,
\left\{\!\!\begin{array}{c}
 c_+-c_-\cos(2\phi) \\
 c_-\sin(2\phi)
\end{array}\!\!\right\} \nonumber\\
&\times\sqrt{\frac{\pi}{\ln 2}\,
\tilde{F}\bigl(
\tfrac{2 z_{\rm R}}{\cal T},
\tfrac{t_0}{\cal T},
\tfrac{{\cal T}_{\rm probe}}{\cal T}
\bigr)} \nonumber\\
&\times\frac{\alpha}{90}\,
\frac{c\,{\cal T}}{\lambda_{\rm probe}}\,
\frac{I({\rm x},{\rm y})}{I_{\rm cr}}\,
{\cal E}_{\rm probe}({\rm x},{\rm y})\,,
\label{eq:pseudofocus}
\end{align}
where $\lambda_{\rm probe}=2\pi/\omega_{\rm probe}$ is the probe wavelength and we reinstated the speed of light in vacuum $c$.
The corresponding signal amplitudes at ${\rm z}\gg{\rm z}_{\rm R, probe}$ can be evaluated with the standard free-space propagation algorithm implemented in LightPipes. To this end, Eq.~\eqref{eq:profiles} is treated as a source in LightPipes: the transverse field profile of the probe ${\cal E}_\mathrm{probe}(\rm{x},\rm{y})$ is taken from the field propagated through LightPipes, and the transverse intensity  distribution of the near-IR laser $I(\rm{x},\rm{y})$ is read from a separate file. Using the versatility of the LightPipes toolbox \cite{LightPipes:1997,LightPipes:2.1.5} these can then be further propagated through optical elements, such as apertures and lenses, up to the detector.

We refer to the expressions in \eqref{eq:pseudofocus} as {\it virtual}, because they do not amount to real physical focus profiles. Upon insertion into Eq.~\eqref{eq:freespaceprop}, they effectively generate the associated physical field distributions at ${\rm z}\gg{\rm z}_{\rm R,probe}$. The fields generated for smaller values of ${\rm z}$ are still to be considered as virtual and differ from the physically realized ones.
For completeness, also note that the linearity of Eq.~\eqref{eq:freespaceprop} in ${\cal E}_{\rm probe}$ renders the ratios ${\cal E}_{{\rm S},p}/{\cal E}_{\rm probe}$ with $p\in\{\parallel,\perp\}$ independent of the probe field amplitude.

\section{Results}\label{sec:Results}

As a benchmark and first illustrative example we consider the collision of an x-ray probe beam featuring an ideal flat-top near-field profile (diameter $d=\,$\SI{200}{\micro\metre} on an ideal absorption-less lens with purely parabolic phase of focal length $f=4\,{\rm m}$ at ${\rm z}=-4\,{\rm m}$) with a pump characterized by an ideal Gaussian focus profile in a counter-propagating geometry. Both beams collide at ${\rm z}=0$ with zero impact parameter.
The relative polarization angle is set to $\phi=\pi/4$, which maximizes the $\perp$-polarized signal component. 
We choose a probe photon energy of $\omega_{\rm probe}=12914\,{\rm eV}$ compatible with the use of a silicon channel-cut polarimeter employing the Si (800) reflection for the detection of the polarization-flipped signal \cite{Marx:2013xwa}.
For the other beam parameters we adopt values available at HED-HIBEF \cite{Ahmadiniaz:2024xob}:
The number of x-ray photons available for probing is $N_{\rm probe}=2\times10^{11}$, and the pump (wavelength \SI{800}{nm}, pulse energy \SI{4.8}{J}) is assumed to be focussed to a FWHM width of \SI{1.3}{\micro\metre}. Finally, the FWHM durations of the pump and probe pulses are ${\cal T}=30\,{\rm fs}$ and ${\cal T}_{\rm probe}=25\,{\rm fs}$.
Somewhat arbitrarily, the detection plane is chosen symmetrically at ${\rm z}=4\,{\rm m}$ which clearly fulfils the far-field condition ${\rm z}\gg{\rm z}_{\mathrm{R,probe}}$. The important point is that this idealized configuration can also be studied analytically \cite{Karbstein:2023}, which facilitates  benchmarking the outcomes of our numerical simulations with the quantum vacuum emission module VIBE put forward in the present work.  
Figure~\ref{fig:2} presents the associated simulation results.  Panels (b) and (c) display maps of the quantum-vacuum signal and the large background of the probe x-rays traversing the counter-propagating high-intensity pump essentially without interaction recoded at ${\rm z}=4\,{\rm m}$.
Lineouts through the peak values of the probe and signal photon distributions at ${\rm y}=0$ are shown in panels (d) and (e), respectively.

Figure~\ref{fig:2} shows that, in line with expectations and analytical estimates, the far-field divergence of the induced quantum vacuum signal is larger than that of the background probe. This behaviour can be traced back to the fact that the transverse extent of the interaction region of the colliding beams around ${\rm z}=0$, and therefore the source of the quantum vacuum signal, is effectively determined by the smaller out of the pump and probe waist sizes. In the present scenario, this is the pump waist. Because the divergence scales inversely with the source size, the signal divergence thus surpasses the probe divergence \cite{Tommasini:2010fb,Karbstein:2015xra}. As a result, the signal exhibits a larger spatial extent at the detection plane. The $\parallel$ and $\perp$ polarized signal components are characterized by the same transverse  distribution. Equation~\eqref{eq:pseudofocus} predicts the associated signal photon numbers to be related as
\begin{equation}
    \frac{N_{{\rm S},\perp}}{N_{{\rm S},\parallel}}=\Bigl(\frac{c_-\sin(2\phi)}{c_+-c_-\cos(2\phi)}\Bigr)^2\ \xrightarrow{\phi=\pi/4}\  \frac{c_-^2}{c_+^2}\approx\frac{9}{121}\,.
\end{equation}

\begin{figure*}[t!]
  \centering
  \includegraphics[width=\textwidth]{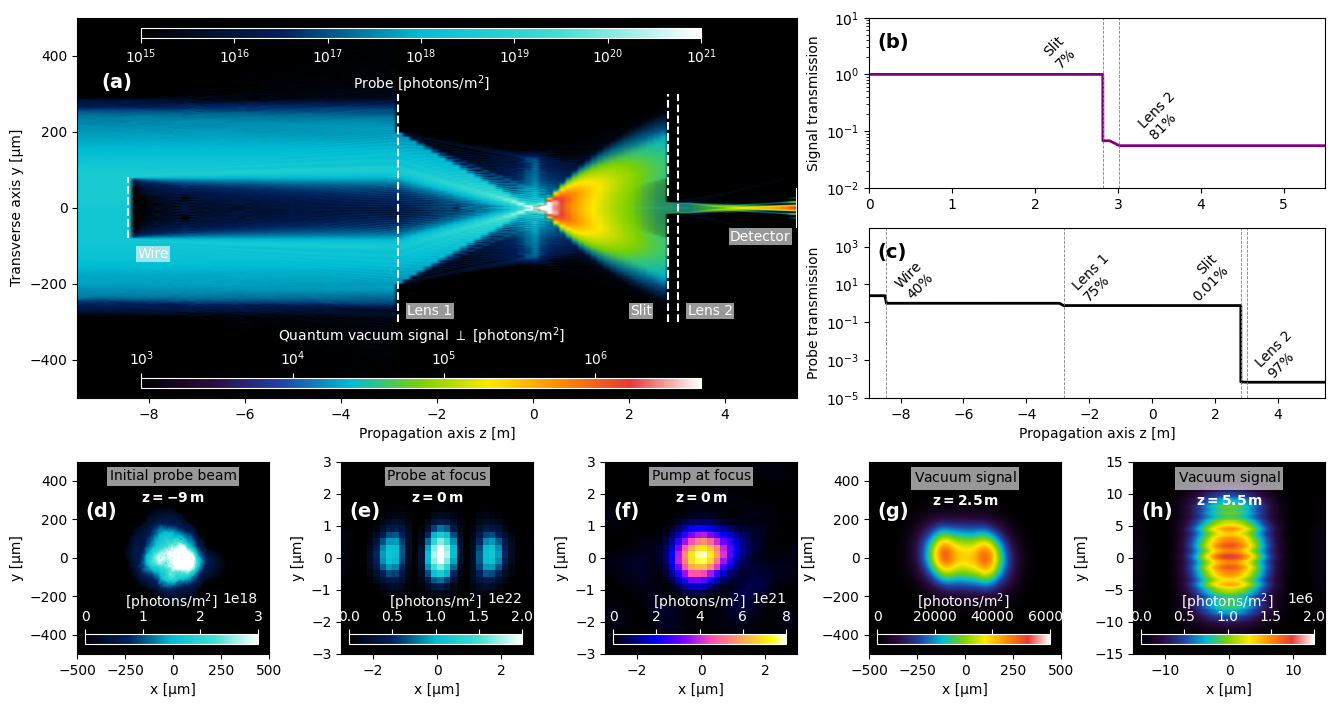}
  \caption{Data-driven simulation of an experimental setup devised to measure the quantum vacuum signal generated in a pump-probe interaction. (a) Side view of the probe (blue) and quantum-vacuum signal (multicolour) intensity maps. Transmission of the (b) probe and (c) signal through the setup. These panels also provide the transmissions of the individual objects inserted in the beam path. Transverse profiles of the (d) initial and (e) focussed probe. (f) Focus profile of the pump. Distribution of the ($\perp$-polarized) quantum vacuum signal (g) right before the slit and (h) in the detector plane.}
  \label{fig:3}
\end{figure*}
A lineout of the quantum vacuum signal is shown in Fig.~\ref{fig:2} (e) together with the results of an analytical calculation \cite{Karbstein:2023}. The excellent agreement in both shape and peak amplitude serves as validation of the numerical implementation in VIBE, and motivates a full demonstration of the capabilities of our approach using experimentally measured beam profiles and realistically modelled optical elements.

To this end, a large-scale simulation was performed that accounts for all the details of a realistic experimental setup devised for the measurement of small quantum vacuum signals. In particular, all beam profiles, apertures, and optical components were modelled with direct input from measured experimental data. This represents a substantial advance beyond previous capabilities. The specific scenario considered corresponds to the dark-field setup \cite{Peatross:1994,Zepf:1998,Karbstein:2020gzg,Karbstein:2022uwf}, which is currently regarded as one of the most promising paths toward a first direct experimental verification of quantum vacuum nonlinearity in macroscopic electromagnetic fields in the laboratory. See Fig.~\ref{fig:3} (a) for an illustration of the implemented setup. Now we employ a probe photon energy of $\omega_{\rm probe}=8766\,{\rm eV}$ compatible with a polarisation-selective beam splitter made of Germanium \cite{Smid:2025wmj}. As in the idealized example discussed previously, we assume $N_{\rm probe}=2\times10^{11}$ and stick to the same parameters ($4.8\,{\rm J}$ in ${\cal T}=30\,{\rm fs}$ at $800\,{\rm nm}$) for the pump as used there. The transverse profile of the initial x-ray probe at ${\rm z}=-9\,{\rm m}$, shown in Fig.~\ref{fig:3} (d), is taken from measurements at the HED instrument of the European XFEL. The probe is partially blocked by a tungsten wire of diameter $d_{\rm wire}=160\,\upmu\mathrm{m}$ before being focussed by a compound refractive lens (CRL) ``Lens~1'' consisting of a stack of two individual beryllium parabolic lenses (radius of curvature $R=50\,\upmu{\rm m}$, wall thickness at the apex $t_{\rm wall}=30\,\upmu{\rm m}$, aperture $A=400\,\upmu{\rm m}$) located further downstream at ${\rm z}=-2.8\,\mathrm{m}$. This results in the interference fringes displayed in Fig.~\ref{fig:3} (e) in the focal plane at ${\rm z}=0$, where the probe is collided with the near-IR pump beam. The transverse distribution of the latter is derived from focus spot measurements for a beam focussed with an f/1 parabola performed at the JETi200 laser system in Jena, Germany, and is shown in Fig.~\ref{fig:3} (f). The associated FWHM width of the focal spot is about $1.3\,\upmu{\rm m}$.
Symmetry of free-space propagation about ${\rm z}=0$ implies that at distances ${\rm z}\gg{\rm z}_{\rm R,probe}$ the transverse profile of the probe background reverts to its incident near-field structure, which is characterized by a central dark-field. By contrast, at ${\rm z}\gg{\rm z}_{\rm R,probe}$ the transverse profile of the quantum vacuum signal is peaked in the dark-field Fig.~\ref{fig:3} (g). This intrinsic directional separation of signal and background provides an efficient discrimination mechanism: a slit formed by the appropriate placement of two tungsten blocks with half-cylindrical inner edges of radius $200\,\upmu{\rm m}$ and a central opening of $30\,\upmu{\rm m}$ at ${\rm z}=2.8\,\mathrm{m}$ then allows to efficiently suppress the large probe background while transmitting the on-axis quantum vacuum signal propagating in forward direction. The signal filtered by the slit is subsequently re-imaged onto the detector at ${\rm z}=5.5\,{\rm m}$ by a second CRL ``Lens~2'' consisting of a stack of four individual beryllium parabolic lenses (radius of curvature $R=50\,\upmu{\rm m}$, wall thickness at the apex $t_{\rm wall}=30\,\upmu{\rm m}$, aperture $A=400\,\upmu{\rm m}$) see Fig.~\ref{fig:3} (h).

Apart from accounting for experimentally realistic pump focus and probe near-field profiles, within our LightPipes/VIBE simulations all optical elements interacting with the probe beam, such as the wire creating the dark-field, the CRLs and the slit, are modelled to reliably reproduce their physical analogues available in experiment.
Within LightPipes, an object of thickness $L({\rm x},{\rm y})$ that depends on the transverse coordinates $\rm x$, $\rm y$ and is characterized by a refractive index $n = 1 - \delta + \rm i\beta$,  with refractive index decrement $\delta$ and absorption index $\beta$, modifies a generic incident transverse beam profile ${\cal E}({\rm x},{\rm y},{\rm z}_0)$ at the longitudinal coordinate ${\rm z}_0$ right in front of the object, via both a phase term 
\begin{equation}
    \Phi({\rm x},{\rm y})=-\frac{2\pi}{\lambda}\delta L({\rm x},{\rm y})
\end{equation}
and a transmission term
\begin{equation}
    T({\rm x},{\rm y})=\exp \left\{-\frac{4\pi}{\lambda}\beta L({\rm x},{\rm y})\right\}\,, 
\end{equation}
such that the resulting transverse field profile directly after the object can be expressed as
\begin{equation}
    {\cal E}({\rm x},{\rm y}{,{\rm z}_0+L}) = {\cal E}_0({\rm x},{\rm y}{,{\rm z}_0})\,\sqrt{T({\rm x},{\rm y})}\; {\rm e}^{{\rm i}\Phi({\rm x},{\rm y})}\,.
\end{equation}
To be specific, in our simulations the wire blocking the incident probe to imprint the dark-field is implemented as a tungsten cylinder with refractive-index parameters $\delta=3.9\times10^{-5}$ and $\beta=2.9\times10^{-6}$ at the considered probe oscillation frequency of $\omega_{\rm probe}=8766\,{\rm eV}$.
Its thickness profile is assumed to be given by $L({\rm x},{\rm y})=2\sqrt{(d_{\rm wire}/2)^2-{\rm x}^{2}}$ within the wire and $L({\rm x},{\rm y})=0$ outside. The CRLs are modelled as stacks of beryllium parabolic lenses ($\delta=4.4\times10^{-6}$, $\beta=1.7\times10^{-9}$ at $\omega_{\rm probe}=8766\,{\rm eV}$) with individual thickness profiles
\begin{equation}
    L({\rm x},{\rm y}) = \frac{{\rm x}^ 2+{\rm y}^2}{R} + t_{\mathrm{wall}}\,,
\end{equation}
The lens aperture $A$ is accounted for by setting to zero the field for $\sqrt{{\rm x}^{2}+{\rm y}^{2}}>A$. 
Importantly, the realistic modelling of the probe beam propagation through the setup also allows to track and assess the transmission of each optical element put in the beam path. See Fig.~\ref{fig:3} (b) for an illustration of the reduction of the photon flux in the probe beam when propagating through the setup, and Fig.~\ref{fig:3} (c) for an analogous plot showing the reduction of the signal photon number from its source to the position of the detector. For a representative detector pixel size of \SI{75}{\micro\metre}, the simulations predict an approximate signal photon yield of $N_{{\rm S},\parallel}={3.56\times 10^{-3}}$ and $N_{{\rm S},\perp}={2.65\times 10^{-4}}$ per optimal pump-probe interaction with zero impact parameter. Our LightPipes/VIBE implementation readily allows for a comparison of these values with the number of background photons $N_{\rm bgr}$ reaching the same detector pixel. For the present, not fully optimized setup the latter is found to be $N_{\rm bgr}=470$. A corresponding optimization is outside the scope of our work reporting on the successful implementation of the vacuum emission module VIBE in LightPipes making such studies possible under experimentally realistic assumptions possible at all. However, note that even for the specific setup studied here the threshold on the polarization purity $\cal P$ of the analyzer required to achieve a signal-to-noise ratio above one for the $\perp$-polarized signal component encoding the vacuum birefringence effect is ${\cal P}<N_{{\rm S},\perp}/N_{\rm bgr}\approx5.6\times10^{-7}$. For x-rays with $\omega_{\rm probe}\sim10\,{\rm keV}$ purities of this order have been demonstrated in experiment \cite{Marx:2013xwa,Schulze:2018,Schmitt:2021,Yu:2023vwe}.

Finally, we emphasize that our numerical simulation toolkit is also quite efficient and lightweight with regards to numerical cost.
A full-scale simulation including all the optical elements introduced above using a transverse resolution of $10000\times10000$ cells and 200 propagation planes in LightPipes requires about five hours on a single 24-core node with 600~GB of RAM. This makes the approach compatible with High Performance Computing resources and even high-end workstations. 
It is worth noting that the high resolution used for the simulations in the present work was primarily chosen for illustration purposes, namely to generate the high-resolution images displayed in Figs.~\ref{fig:2} and \ref{fig:3}. Much lighter simulations using, e.g. $4000\times4000$ cells and only six propagation planes in LightPipes, yield essentially the same physics predictions, while reducing the runtime to approximately one hour for given computational resources. This efficiency makes the LightPipes/VIBE simulation toolkit particularly well suited for explorative parameter scans, geometry optimization, and rapid analysis workflows.

\section{Conclusions and Outlook}\label{sec:Concls+Outlook}

In the present work we have demonstrated and put forward the {\it Vacuum Interaction Birefringence Explorer} (VIBE), a numerical simulation tool facilitating the reliable study of quantum vacuum signals in experimentally realistic setups within the framework of the diffractive beam propagation toolbox LightPipes.
VIBE heavily benefits from the versatility of LightPipes and, in its current version, specifically enables quantitatively accurate studies of the quantum vacuum signals generated in the collision of counter-propagating laser beams in pump-probe type experiments with a weakly-focussed probe at unprecedented detail.
The latter condition is in particular met by forthcoming vacuum birefringence experiments using an x-ray free-electron laser as probe.

Simulations with LightPipes/VIBE allow to account for the experimentally realistic focus profiles of the colliding laser beams, and consistently propagate both the probe background and the induced quantum vacuum signal through a complete model of the actual experimental setup. 
We are convinced that these capabilities are indispensable for the accurate planing, implementation and subsequent analysis of this type of experiments.
The moderate numerical costs of a single simulation run, will enable surveying large parameter spaces and studying the fate of the quantum vacuum signals and associated probe background for various different initial conditions. In combination with a Monte Carlo algorithm that statistically samples the parameter fluctuations inherent to any real experimental implementation, this will allow to reliably account for beam misalignment and jitter effects and thus provide important insights beyond state-of-the-art simulation capabilities.

Finally, we note that there are several natural ways to advance VIBE in the future, ranging from extensions to other longitudinal beam profiles and collision geometries, via adaptations to scenarios involving more than one pump beam, towards eventually the combination with a numerical Maxwell solver allowing to determine the quantum vacuum signals from first principles.

The vacuum emission module \href{https://github.com/amatheron/VIBE}{VIBE} for LightPipes is available in open source and should be cited as: A. Matheron \textit{et al.}, \href{https://doi.org/10.5281/zenodo.17979735}{doi:10.5281/zenodo.17979735} (2025).

\acknowledgements

This work has been funded also by the Deutsche Forschungsgemeinschaft (DFG) under Grants No. 416607684, No. 416702141, and No. 416708866 within the Research Unit FOR2783/2.

\bibliographystyle{abbrv}

\end{document}